\documentstyle[12pt]{article}  
\newcommand{\beq}{\begin{equation}} \newcommand{\eeq}{\end{equation}}
\newcommand{\bra}{\begin{array}} \newcommand{\era}{\end{array}}

\newcommand{\ot}{\otimes}

\begin{document}
\begin{center}
{\Large\bf A Generalized Jaynes-Cummings Model:
 Nonlinear dynamical superalgebra $u(1/1)$ and Supercoherent states}
\vskip1.2truecm
{\bf M. Daoud} 
\footnote{Permanent address: University of Ibn Zohr, L.P.M.C., Departement of Physics, B.P. 28/S, Agadir, Morocco.}
and {\bf J. Douari}\\
\vskip.5cm
The Abdus Salam ICTP - Strada Costiera 11, 34100 Trieste, Italy.\\
\vspace*{0.5cm} 
\end{center}
\hoffset=-1cm\textwidth=20cm\pagestyle{empty} \vspace*{0.5cm}
\begin{abstract}
\hspace{.3cm}The generalization of the Jaynes-Cummings (GJC)
Model is proposed. In this model,
the electromagnetic radiation is described by a Hamiltonian generalizing the
harmonic oscillator to take into account some nonlinear effects which can
occurs in the experimental situations. The dynamical superalgebra and
supercoherent states of the related model are explicitly constructed. A
relevant quantities (total number of particles, energy and atomic inversion)
are computed.
\end{abstract}
\vskip1.2truecm
\section{Introduction}
\hspace{.3cm}The Jaynes-Cummings (JC) model \cite{1} which is extensively used in
quantum  optics describes, in its simplest version, the interaction of a
cavity mode with two-level system \beq H_{JC}=\omega(a^+ a^- +
\frac{1}{2})\sigma_0 +\frac{\omega_0 }{2}\sigma_3 +\kappa (a^+ \sigma_- +a^-
\sigma_+ ), \eeq where $a^+ $ and $a^- $ are the photon creation and
annihilation operators, $\sigma_\pm =\frac{1}{2}(\sigma_1 \pm\sigma_2 )$,
with $\sigma_1 $, $\sigma_2 $ and $\sigma_3 $ are the Pauli matrices and
$\sigma_0 $ is the identity matrix. Moreover, $\kappa$ is a coupling
constant, $\omega$ is the radiatif field mode frequency and $\omega_0 $ the atomic
frequency. The interest of this model, its solvability and its
applications, has long been discussed \cite{1,2,3}. Over the last two
decades, there has been intensive study [4 and references quoted therein]
on the solvable Jaynes-Cummings model and its various extensions, such as
intensity depending coupling constants, two photons or multiphoton
transitions and two or three cavity modes for three-level atoms. These
models have found their applications in laser trapping and cooling atoms
\cite{5} and quantum nondemolition measurements \cite{6}. Furthermore, the
Jaynes-Cummings model constitutes now the basis for a vast array of the
current experiments on fundations of quantum mechanics involved entangled
states [7, and references therein]. In other hand, the supergroup
theoretical approach to Jaynes-Cummings model has opened the way to
relate the exact solvability of this model and representation theory of
superalgebras. Indeed the Hamiltonian $H_{JC}$ is an element of the
$u(1/1)$ superalgebra \cite{8}. In the absence of coupling ($\kappa=0$)
and for exact resonance ($\omega=\omega_0 $), the $u(1/1)$ dynamical superalgebra
reduces to $sl(1/1)$ one and the JC model coincides with the
supersymmetric harmonic oscillator. More recently, the investigations of
a class of shape-invariant bound state problem, which represents
two-level system, leads to generalized Jaynes-Cummings model
\cite{9,10,11}. In the case of the simplest shape-invariant system,
namely the harmonic oscillator, the generalized Jaynes-Cummings model
reduces to standard one.\\

In this paper we shall address the generalization and quantum characteristics
of the Jaynes-Cummings model. Besides the eigenvalues and eigenvectors,
we give the supercoherent states of the related model. It is found that the
generalized Jaynes-Cummings model is governed   by a nonlinear
superalgebra $u(1/1)$ which reduces to well known $u(1/1)$ occurring in
the standard  (JC) model \cite{12,13}. We compute the total number of
photons and the energy. We find that the atomic inversion exhibits  Rabi
oscillations.  \\

The paper is organized as follows: In Section 2, we introduce the generalized
supersymmetric quantum oscillator and we construct the corresponding
supercoherent states. Exact spectrum of the generalized Jaynes-Cummings
model is given in section 3. Section 4 is devoted to nonlinear dynamical
superalgebra  $u(1/1)$ of the (GJC) model which is useful to construct the
supercoherent states adapted to our model (Section 5). Using  the latter set of
super-states, we compute in section 6 some relevant physical quantities.
The last section concerns the conclusion of this work.\\

 \section{Generalized Supersymmetric Quantum Oscillators}
\hspace*{.3cm}We begin by introducing the generalized supersymmetric
quantum oscillators. Let us Consider a Hamiltonian $H$ with a discrete
spectrum which is bound below and has been adjusted so that $H\ge 0$. We
assume that the eigenstates of $H$ are non-degenerate. The eigenstates
$\vert\Psi_n\rangle$ of $H$ are orthonormal vectors and they satisfy
\beq
H\vert
\Psi_n\rangle=e_n \vert \Psi_n\rangle.
\eeq
In a general setting, we also assume that the energies $e_0 ,e_1 , e_2 ,...$
are positive and verify $e_{n+1}>e_n $. The ground state energy is $e_0
=0$. We define the creation and the annihilation operators $A^+ $ and $A^-
$, respectively, such that the Hamiltonian can be factorized as \beq H=A^+
A^- . \eeq   The action of the operators $A^+ $ and $A^- $ on the states
$\vert \Psi_n \rangle $  are given by
\beq
\bra{rcl}
A^+ \vert \Psi_n\rangle &=&(e_{n+1})^{\frac{1}{2}}\vert \Psi_{n+1}\rangle
\\ A^- \vert \Psi_n\rangle &=&(e_{n})^{\frac{1}{2}}\vert \Psi_{n-1}\rangle
\era
\eeq
implemented by the action of $A^- $ on the gound state $\vert \Psi_{0}\rangle$  \beq A^- \vert \Psi_{0}\rangle=0.
\eeq
The commutator of $A^+ $ and $A^- $ is defined by \beq
\lbrack A^- ,A^+ \rbrack =G(N), \eeq
where the operator $G(N)$ is defined through his action on $\vert \Psi_n \rangle$
\beq G(N) \vert \Psi_n\rangle
=(e_{n+1}-e_{n}) \vert \Psi_n\rangle .
\eeq
We define the number operator $N$ as \beq N\vert \Psi_n\rangle =n\vert
\Psi_n\rangle ,
\eeq
$N$ is in general different from the product $A^+ A^- $
(=H). We can see that the number operator satisfy \beq \bra{rcl} \lbrack A^+
,N \rbrack&=& A^+  \\
\lbrack A^- ,N \rbrack&=& -A^- . \era \eeq
Here, we consider two generalized oscillators systems which has extensively
studied in the literature. The first concerns the so-called generalized
deformed oscillator \cite{14} and the second one is the $x^4$-anharmonic
oscillator \cite{15}. The physical interests of this two systems have been
extensively enumerated [see the references quoted in \cite{14,15}]. Here, we
recall their eigenstates and eigenvalues to construct the supersymmetric
generalized quantum oscillator and the corresponding supercoherent states.\\

To introduce the generalized deformed oscillator, the procedure of \cite{14}
requires the existence of a map  from the usual harmonic oscillator algebra
generated by annihilation and creation operators $a^-$and $a^+$  satisfying
the standard canonical commutation relations, to the new one generated by
$A^-$ and $A^+$
 \beq
 A^- = a^- f(N)  \:\:\:\,\:\:\:\ A^+ =  f(N) a^+
 \eeq
$N$ being the number operator $N = a^+a^-$ and the function $f$ is given
by
 \beq
 f(N) = N+m
 \eeq
The Hamiltonian of the obtained generalized harmonic oscillator is then given
by
 \beq
 H = A^+A^- = N(N+m)
 \eeq
 with eigenvalues
 \beq
 e_n = n(n+m)
 \eeq
The Fock states 
 $\vert \Psi_n\rangle\equiv\vert
n,m\rangle$ are labelled by the integers $m$ and $n=0,1,2,...$.

It is clear that the operator $G(N)$ (7), in this case is given by
\beq
 G(N) = 2N+m+1.
 \eeq
Then, the  operators $A^+, A^-$ and $G(N)$ satisfy the relations (6) and
(9). Note that other choices of the function $f$ are possible. We remark  also
that when $f(N)= 1$, we have the ordinary harmonic oscillator.

The other nonlinear oscillator that we consider is the $x^4$-anharmonic
oscillator. The Hamiltonian, describing this  system, is 
\beq
H = a^+a^-+\frac{\epsilon}{4}(a^-+a^+)^4-\delta 
\eeq
where $ a^+$ and $a^-$ are the creation and annihilation operators for the
harmonic oscillator. The quantity $\delta$ is given by
\beq
\delta = \frac{3}{4}\epsilon - \frac{21}{8}{\epsilon}^2
\eeq 
which vanishes when $\epsilon = 0$ and $H$ reduces in this limit to the
standard harmonic  oscillator Hamiltonian.  The Hamiltonian $H$ can be
factorized in the following form \cite{15}
 \beq
 H = A^+A^-
 \eeq
in terms of $A^+$ and $A^-$ which are expressed as some functions of
$a^+$ and $a^-$ (for the   expressions of these functions see\cite{15}). The
energy levels are given by 
 \beq \bra{rl} e_n
=n+\frac{3}{2}\epsilon(n^2 +n),& n=0,1,2,... \era \eeq The positive parameter
$\epsilon$ can  be seen as one taking into account some non linearity of the
radiatif field arising from some perturbative effects occurring in experimental
situations. Note that we keep terms only up to $\epsilon$ which  is the
standard first-order perturbation result.  The Hilbert space of this system is
easily constructed  in the same way as the standard harmonic oscillator. It is
spanned by the states $\vert n,\epsilon\rangle$, $n=0,1,2,...$ which is
generated by the action of $A^+ $ on the ground state $\vert
0,\epsilon\rangle$. The operator $G(N)$ is
\beq
G(N) = 1+3{\epsilon}(n+1)
\eeq
Here, again one can verify that the relations (6) and (9) are satisfied by the creation and annihilation
operators corresponding to the $x^4$-anharmonic system.

In supersymmetric quantum mechanics, one consider the so-called
supersymmetric Hamiltonian which is defined by \beq \bra{rcl}
H_{susy}=\pmatrix{A^- A^+ &0\cr 0&A^+ A^- \cr}=\pmatrix{H_{+} &0\cr 0&H_{-}
\cr}, \era \eeq where $H_{+}=A^- A^+$ and $H_{-}=A^+ A^- =H$ are the
so-called supersymmetric partner Hamiltonians.

Working in the Hilbert
space \beq h=h_b \ot h_f =\left\{\vert \Psi_n ,-\rangle=\pmatrix{0\cr \vert
\Psi_n\rangle\cr}, \vert \Psi_n ,+\rangle=\pmatrix{\vert \Psi_n\rangle\cr
0\cr}; n=0,1,2,...\right\}, \eeq the eigenstates are \beq \bra{lll} \vert
\phi_0\rangle=\pmatrix{0\cr \vert \Psi_0\rangle\cr},&\\ \vert
\phi_{n>0}\rangle=c^{+}_{n}\pmatrix{\vert \Psi_{n-1}\rangle\cr
0\cr}+c^{-}_{n}\pmatrix{0\cr \vert \Psi_n\rangle\cr},& |c^{+}_{n}|^2
+|c^{-}_{n}|^2 =1, \era \eeq
with the energies $E_0 =0$ and $E_{n>0}=e_n $. Because, we are interested
by generalized quantum oscillator, the states $\vert \Psi_n\rangle$ are $\vert
n,m\rangle$ for the generalized deformed oscillator and $\vert \Psi_n\rangle$
are $\vert n,\epsilon\rangle$ for the $x^4$-anharmonic oscillator. As we
have mentioned previously these two quantum systems will be used to
extend the JC model and we will compute some relevant physical  quantities,
like the mean values of the total number operators, the energy and the atomic
inversion, over the coherent states of the (GJC) model. The latter will be
obtained from the supercoherent states corresponding the generalized
supersymmetric oscillator. So  for $H_{susy}$, we consider the
supercoherent states (linear combination of the fondamental coherent states)
\beq
\vert
z,\beta\rangle=\cos\frac{\theta}{2}\pmatrix{0\cr \vert z\rangle\cr}+\sin\frac{\theta}{2}e^{i\Phi}\pmatrix{\vert z\rangle\cr
0\cr},
\eeq
with $\beta=\frac{\theta}{2}e^{i\Phi}$ and $\vert z\rangle$
 are the coherent states, corresponding to $H_-$, defined by \cite{16,17,18,19}
 \beq \vert z\rangle=\aleph
(|z|)\sum\limits_{n=0}^{\infty}\frac{z^n }{(e(n))^{\frac{1}{2}}}\vert \Psi_n\rangle,
\eeq
where $e(n)=e_1 ...e_n $ for $n=1,2,...$ and we set $e(0)=1$. The
normalization constant $\aleph (|z|)$ is calculated from the normalization
condition $\langle z\vert z\rangle=1$ and it is given by
\beq
(\aleph(|z|))^{-2}=\sum\limits_{n=0}^{\infty}\frac{|z|^{2n}}{e(n)}.
\eeq
Let us mention that the coherent states for $x^4$-anharmonic oscillator has
been studied in \cite{19}. They are given by
\beq
\vert z\rangle=a(|z|)\sum\limits_{n=0}^{\infty}(\frac{2^n }{(3\epsilon)^n \Gamma(n+1)\Gamma(n+2+\frac{2}{3\epsilon})})^{\frac{1}{2}}z^n \vert n,\epsilon\rangle,
\eeq
with
\beq
a(|z|)=\frac{(\Gamma(2+\frac{2}{3\epsilon}))^{\frac{1}{2}}}{(_{0}F_{1}(2+\frac{2}{3\epsilon},\frac{2}{3\epsilon}|z|^2))^{\frac{1}{2}}}.
\eeq

For the deformed generalized harmonic oscillator, we construct the coherent
states in the same way that one which gives Barut-Girardello coherent states
of the $su(1,1)$ algebra \cite{20}.  Note that the algebra generated by
$\left\{A^+ ,A^- ,G(N)=2N+m+1\right\}$ is   isomorphic to $su(1,1)\sim
sl(2,{\bf(R)\rm})\sim so(2,1)$. Indeed, the creation $A^+$ and annihilation
$A^- $    operators satisfy the following commutation relations
\beq \bra{rl}
\lbrack A^- ,A^+ \rbrack=G(N),&
\lbrack A^{\pm} ,G(N) \rbrack=\mp A^{\pm}.
\era \eeq

A more familiar basis for $su(1,1)$ algebra is given by
\beq \bra{rcl} J_-
=\frac{1}{\sqrt{2}} A^- ,&J_+ =\frac{1}{\sqrt{2}}A^+ ,& J_{12}=\frac{1}{2} G(N),  
\era \eeq
with the following commutation relations
\beq \bra{rl}
\lbrack J_- ,J_+ \rbrack=J_{12},& \lbrack J_{\pm} ,J_{12} \rbrack=\mp
J_{\pm}.  \era \eeq
Barut and Girardello introduced the $su(1,1)$ coherent states as
eigenvectors of $J_-$ \cite{20}
\beq \bra{rcl} J_- \vert z\rangle &=&z\vert
z\rangle \\
\vert z\rangle &=&b(|z|)\sum\limits_{n=0}^{\infty}\frac{(\sqrt{2}z)^n
}{(n!\Gamma(n+m+1))^{\frac{1}{2}}}\vert n,m\rangle . \era \eeq
The normalization constant is given by
\beq
b(|z|)=\frac{(\Gamma(m+1))^{\frac{1}{2}}}{(_{0}F_{1}(m+1,2|z|^2 ))^{\frac{1}{2}}}.
\eeq
Where the $_{0}F_{1}(m+1,2|z|^2 )$ is the hypergeometric function. The
coherent states (26) ( for the $x^4$-anharmonic oscillator) and (31) (fot the
generalized  deformed oscillator) can be obtained simply from equation (24)
by replacing in the expressions of $e(n)$ the energies by their  
corresponding values for each considered system. Note that the coherent
states (26) and (31) will be useful to build up ones of (GJC) model (see
section 5). Remark also that the resolution to identity of such states has not
been discussed here. However, the measures by respect with the previous
sets of coherent states are over-completes can be computed in a very easy
way following the approach developed in \cite{13,17}.\\  
 
\section{Eigenstates and Eigenvalues}
\hspace{.6cm}To generalize the JC model, let us consider the Hamiltonian
\beq
H_{GJC}=\frac{1}{2}\omega \lbrace A^- ,A^+ \rbrace +\frac{1}{2}\omega _0
\lbrack A^- ,A^+ \rbrack\sigma_3 +\kappa(A^{+}\sigma_- + A^{-}\sigma_+ ).
\eeq
This last expression generalizes the ordinary JC Hamiltonian. In fact, when
the creation and the annihilation operators are those associated with the
harmonic oscillator, the above Hamiltonian reduces to the well known JC
model (Eq.(1)). We note also that the Hamiltonian $H_{GJC}$ is
supersymmetric when $\omega =\omega _0 $ (exact resonance) and
$\kappa=0$ (absence of coupling)
\beq
H_{GJC}(\kappa=0,\omega =\omega
_0 )=\lbrace Q^- ,Q^+  \rbrace ,
\eeq
where the supercharges operators are defined by
\beq \bra{rl} Q^-
=i\sqrt{\omega }a^+
\sigma_- ,& Q^+ =-i\sqrt{\omega }a^- \sigma_+ , \era
\eeq
and satisfy the relations
\beq \bra{rcl} (Q^- )^2 =0 ,& (Q^+ )^2 =0 ,& \lbrack Q^\pm ,H
\rbrack=0. \era
\eeq
We note that the generalized Hamiltonian of JC model (33) is different from
ones considered in the references \cite{21,22,23}. The main purpose of this
section is to show that generalized JC model can be solved analytically. The
solutions become ones corresponding to JC model when $A^+ =a^+ $ and
$A^- =a^- $ (i.e. harmonic oscillator).\\

The diagonalization of the Hamiltonian $H_{GJC}$ is easily carried out
(in the same way that the standard JC model) and leads to the eigenstates
\beq \bra{rcl} \vert E_{0}^{-}\rangle&=&\vert \Psi_0 , -\rangle\\ \vert
E_{n}^{+}\rangle&=&\frac{1}{P(n+1)}\lbrack S(n+1)\vert \Psi_n , +\rangle
-Q(n+1)\vert \Psi_{n+1} , -\rangle\rbrack\\ \vert
E_{n+1}^{-}\rangle&=&\frac{1}{P(n+1)}\lbrack Q(n+1)\vert \Psi_n ,
+\rangle +S(n+1)\vert \Psi_{n+1} , -\rangle\rbrack , \era \eeq where 
\beq
\bra{rcl}
Q(n+1)&=&\kappa(e_{n+1})^{\frac{1}{2}}\\
S(n+1)&=&\kappa(e_{n+1}+\frac{\Delta_-}{2\kappa^2}(\frac{e_{n+2}-e_{n}}{2})^2)^{\frac{1}{2}}+
\frac{\Delta_-}{2}\frac{e_{n+2}-e_{n}}{2},\\
P(n)&=&((S(n))^2+(Q(n))^2)^{\frac{1}{2}}\era \eeq
with $\Delta_- =\omega -\omega _0 $.\\

The corresponding eigenvalues are given by
\beq \bra{rcl}
E_{n}^{+}&=&\frac{\Delta_+}{2}e_{n+1}+\frac{\Delta_-}{4}(e_{n+2}+e_{n})-(\kappa^2
e_{n+1}+\frac{\Delta_{-}^{2}}{4}(\frac{e_{n+2}-e_{n}}{2})^{2})^{\frac{1}{2}}\\
E_{n+1}^{-}&=&\frac{\Delta_+}{2}e_{n+1}+\frac{\Delta_-}{4}(e_{n+2}+e_{n})+(\kappa^2
e_{n+1}+\frac{\Delta_{-}^{2}}{4}(\frac{e_{n+2}-e_{n}}{2})^{2})^{\frac{1}{2}}, \era \eeq
where $\Delta_+ =\omega +\omega _0 .$

Interesting particular cases arise from the former results. We start with the
exact resonance $(\Delta_- =0)$. In this case, the Hamiltonian $H_{GJC}$
takes the form
\beq
H_{GJC}(\Delta_{-}=0)=\frac{1}{2}\omega \lbrace A^-
,A^+ \rbrace +
\frac{1}{2}\omega \lbrack A^- ,A^+ \rbrack \sigma_3 + \kappa(A^+ \sigma_- +A^-
\sigma_+ ).
\eeq
The eigenstates of the resonant generalized JC model are then
\beq \bra{rcl} \vert E_{n}^{+}\rangle&=&\frac{1}{\sqrt{2}}\lbrack
\vert \Psi_n , +\rangle -\vert \Psi_{n+1} , -\rangle\rbrack\\ \vert
E_{n+1}^{-}\rangle&=&\frac{1}{\sqrt{2}}\lbrack \vert \Psi_n , +\rangle
+\vert \Psi_{n+1} , -\rangle\rbrack . \era
\eeq
and the corresponding eigenvalues are
\beq \bra{rcl} E_{n}^{+}&=&\omega 
e_{n+1}-\kappa\sqrt{e_{n+1}}\\ E_{n+1}^{-}&=&\omega 
e_{n+1}+\kappa\sqrt{e_{n+1}}. \era
\eeq
The resonant generalized JC model is reduced to supersymmetric
Hamiltonian when the coupling constant $\kappa$ vanishes
\beq
H_{susy}=H_{GJC}(\Delta_{-}=0=\kappa)=\omega
\pmatrix{A^- A^+ &0\cr 0&A^+ A^-
\cr}.
\eeq

Finally, we remark that when $e_n =n$ (spectrum of the quantized radiatif
field), the eigenstates and eigenvalues (37) and (39) coincide with ones
corresponding to standard JC model \cite{12} (see also \cite{13}).
\section{Dynamical Superalgebra}
\hspace{.6cm}Like standard JC model \cite{12}, the generalized JC model can be written as a linear combination
of two even operators $N_1$ and $N_2$ and two odd operators
$Q^-$ and $Q^+$; \beq
H_{GJC}=\frac{\Delta_+}{2}N_{1}-\frac{\Delta_-}{2}N_{2}+\frac{i\kappa}{\sqrt{\omega }}(Q^{+}-Q^{-}),
\eeq where \beq \bra{rcl} N_{1}=\pmatrix{A^- A^+ & 0\cr 0&A^+ A^-
\cr},& N_{2}=\pmatrix{-A^+ A^- & 0\cr 0&-A^- A^+ \cr}\\ \\
Q^{-}=i\sqrt{\omega }A^+ \sigma _- ,& Q^{+}=-i\sqrt{\omega }A^- \sigma _+ . \era \eeq
These operators satisfy the following relations
\beq
\bra{llll}
\lbrack N_{1}, N_{2}\rbrack=0,&
\lbrack N_{1},Q^{-}\rbrack=\lbrack N_{1},Q^{+}\rbrack=0,&\\
\lbrack N_{2},Q^{+}\rbrack=(Q^{+}g_{N}+g_{N}Q^{+}),& 
\lbrack N_{2},Q^{-}\rbrack=-(Q^{-}g_{N}+g_{N}Q^{-}),& \\
\lbrace
Q^{-},Q^{+}\rbrace=\omega N_{1},& (Q^{-})^2 = (Q^{+})^2 =0,
\era
\eeq
where the even operator $g_{N}$ is defined by
\beq g_{N}=G(N)\sigma _0 .
\eeq
The algebra generated by $\left\{ N_{1}, N_{2}, Q^{-},
Q^{+}\right\}$ can be seen as a non linear (or deformed) version of
the superalgebra $u(1/1)$. Indeed, when $G(N)=1$ (a situation which
occurs in the ordinary JC model), we have the $u(1/1)$ superalgebra. In
the situation, in which we are in presence of an exact resonance
$w=w_0 $, the generalized JC model is written as
\beq
H_{GJC}=\omega N_{1}+\frac{i\kappa}{\sqrt{\omega }}(Q^{+}-Q^{-}),
\eeq
in terms of the operators $N_{1}$, $Q^{+}$ and $Q^{-}$ which satisfy the
structural relations of the superalgebra $sl(1/1)$. It is also easy to see
that for $\omega =\omega _0 $ and $\kappa=0$, we have
\beq
\bra{lll}
H_{GJC}=\omega N_{1}=\lbrace Q^{+},Q^{-}\rbrace,&\\ \lbrack
Q^{\pm},H_{GJC}\rbrack=0,& (Q^{-})^2 = (Q^{+})^2 =0.
\era
\eeq
\section{Supercoherent states for $H_{GJC}$}
\hspace{.6cm}To construct the supercoherent states for $H_{GJC}$, we start by looking
for an unitary operator $U$ which connects the Hamiltonians $H_D$
(diagonal in the fondamental space of states $\left\{\vert \Psi_n
,+\rangle, \vert \Psi_n
,-\rangle\right\}$) and $H_{GJC}$
\beq
H_D =U^+ H_{GJC}U=\pmatrix{H_+ &0\cr
0&H_- \cr},
\eeq
where
\beq
\bra{rcl}
H_+ &=&\frac{\Delta_+ }{2}g(N+1)+\frac{\Delta_-
}{4}(g(N+2)+g(N))-(\kappa^2 g(N+1)+\frac{\Delta_- }{4}(\frac{g(N+2)-g(N)}{2})^2 )^{\frac{1}{2}}\\
H_- &=&\frac{\Delta_+ }{2}g(N)+\frac{\Delta_-
}{4}(g(N+1)+g(N-1))+(\kappa^2 g(N)+\frac{\Delta_- }{4}(\frac{g(N+2)-g(N)}{2})^2 )^{\frac{1}{2}}.
\era
\eeq
The operators $g(N+k)$ ($k=-1,0,1,2$) are defined by
\beq
\bra{rcl}
g(N+k)\vert \Psi_n \rangle&=& e_{n+k} \vert \Psi_n \rangle .
\era
\eeq
The operator $U$ takes the following form
\beq
U=\pmatrix{\frac{1}{P(N+1)}S(N+1)&\frac{\kappa}{P(N+1)}A^- \cr
-A^+ \frac{\kappa}{P(N+1)}& \frac{1}{P(N)}S(N) \cr},
\eeq
where
\beq
S(N)=\frac{\Delta_-
}{2}\frac{g(N+1)-g(N-1)}{2}+\kappa(g(N)+\frac{\Delta_- }{2\kappa}(\frac{g(N+1)-g(N-1)}{2}))^{\frac{1}{2}},
\eeq
and
\beq
P(N)=(S(N)^2 +\kappa^2 g(N))^{\frac{1}{2}}.
\eeq
The operator $U$ can be written as follows
\beq
U=e^{-Z},
\eeq
with
\beq
Z=A^+ h(N+1)\sigma_- -h(N+1)A^- \sigma_+ .
\eeq
Developing $e^{-Z}$ and using the following identities of the Hermitian
skew operator $Z$,
\beq
Z^{2p}=(-1)^p (h(N+1))^{2p}(g(N+1))^{p}\sigma_+ \sigma_- + (-1)^p (h(N))^{2p}(g(N))^{p}\sigma_- \sigma_+
\eeq
and
\beq
Z^{2p+1}=(-1)^{p+1}(h(N+1))^{2p+1}(g(N+1))^{p}A^- \sigma_+ + (-1)^p A^+ (h(N+1))^{2p+1}(g(N+1))^{p}\sigma_- .
\eeq
It becomes that $U$ is given
\beq
U=\pmatrix{\cos(h(N+1)(g(N+1))^{\frac{1}{2}})&\frac{-1}{(g(N+1))^{\frac{1}{2}}}\sin(h(N+1)\frac{1}{(g(N+1))^{\frac{1}{2}}})
A^- \cr
A^+ \frac{1}{(g(N+1))^{\frac{1}{2}}}\sin(h(N+1)(g(N+1))^{\frac{1}{2}})&\cos(h(N)(g(N))^{\frac{1}{2}}) \cr},
\eeq
where $h(N)=\frac{-1}{\sqrt{g(N)}}\arctan(\kappa\frac{\sqrt{g(N)}}{S(N)})$.
Of course, for $g(N)=N$ (standard JC model), we recover the results obtained in
\cite{12}.\\

Using the unitary operator $U$, we introduce the coherent states for
generalized JC Hamiltonian as follows
\beq
\bra{rcl}
\vert z,\beta\rangle_{GJC}&=&U\vert z,\beta\rangle\\
&=&\aleph (|z|)\sum\limits _ {n=0} ^ {\infty}\frac {z^n}{\sqrt {e(n)}} (\sin\frac {\theta}{2} e ^ {i\Phi}\vert
E^{+}_{n}\rangle +\cos\frac {\theta} {2}\vert E ^ {-} _ {n}\rangle).
\era
\eeq
To compute some relevant physical quantities of the generalized JC model,
we consider the time evolution of states (61)
\beq
\vert z,\beta ,t\rangle_{GJC}=e^{-itH_{GJC}}\vert z,\beta\rangle.
\eeq
We get
\beq
\vert z,\beta ,t\rangle _{GJC}
=\aleph (|z|)\sum\limits_{n=0}^{\infty}\frac {z^n}{(e(n))^{\frac{1}{2}}}(\sin\frac
{\theta}{2}e^{i\Phi}e^{-itE^{+}_{n}} \vert E^{+}_{n} \rangle + \cos \frac
{\theta}{2} e^{-itE^{-}_{n}} \vert E^{-}_{n} \rangle ).
\eeq
Using the latter states, we will compute the average value of the total value of photons,
energy and atomic inversion.
\section{Total number of photons, energy and atomic inversion}
\subsection{Total number of particules}
\hspace{.6cm}The mean values of the operator $N$ over the states (63) is
given by
\beq
\langle N\rangle = \aleph (|z|)^2
\sum\limits_{n=0}^{\infty}\frac{|z|^{2n}}{e(n)}(\sin^2
\frac {\theta}{2}\langle E^{+}_{n}\vert\hat N\vert E^{+}_{n}\rangle
+\cos^2 \frac{\theta}{2}\langle E^{-}_{n}\vert\hat N\vert
E^{-}_{n}\rangle).
\eeq
A direct computation of matrix elements occuring in the last expression
gives
\beq
\langle N\rangle = \aleph(|z|)^{2}
\sum\limits_{n=0}^{\infty}\frac{|z|^{2n}}{e(n)}(\sin^2
\frac{\theta}{2}e_{n+1}+\cos^2 \frac{\theta}{2}e_n ).
\eeq
where the energies $e_n$ are given (13) (resp.(18)) for the generalized deformed oscillator 
(resp. for the $x^4$-anharmonic system). 
\subsection{Energy}
\hspace{.6cm}The energy in coherent states (63) is, simply, given by
\beq
\langle H_{GJC}\rangle = \aleph(|z|)^2 
\sum\limits_{n=0}^{\infty}\frac{|z|^{2n}}{e(n)}(\sin^{2}
\frac{\theta}{2}E^{+}_{n}+\cos^2 \frac{\theta}{2}E^{-}_{n}).
\eeq
The ${E_n}^+$ and ${E_n}^-$ are given by Eqs (39).
\subsection{Atomic inversion}
\hspace{.6cm}First, if we start with supercoherent states (61), we see that the average value of the third component of
the spin is time independent. However, as it is well known, if the radiatif field is prepared in a coherent state, 
the atomic inversion consists of Rabi oscillations. The temporal dependence of $\langle\sigma_{3}\rangle$ appears when
we use the generalized supercoherent states (63). Indeed, we get
\beq
\langle\sigma_{3}\rangle =
\langle\sigma_{3}\rangle_{++}+\langle\sigma_{3}\rangle
_{--}+\langle\sigma_{3}\rangle _{+-}+\langle\sigma_{3}\rangle_{-+},
\eeq
with
\beq
\bra{rcl}
\langle\sigma_{3}\rangle_{++}&=&\frac{1}{2}(1-\cos\theta)\aleph (|z|)^{2} 
\sum\limits_{n=0}^{\infty}\frac{|z|^{2n}}{e(n)}q(n+1)\\
\langle\sigma_{3}\rangle_{--}&=&-\frac{1}{2}(1+\cos\theta)\aleph (|z|)^{2} 
\sum\limits_{n=0}^{\infty}\frac{|z|^{2n}}{e(n)}q(n)\\
\langle\sigma_{3}\rangle_{+-}&=&\frac{1}{2}\sin\theta\aleph (|z|)^{2} 
\sum\limits_{n=0}^{\infty}\frac{|z|^{2n+1}}{e(n)}e^{it(2\kappa
s(n+1)+\Phi-\alpha)}\frac{1}{s(n+1)}\\
\langle\sigma_{3}\rangle_{-+}&=&\frac{1}{2}\sin\theta\aleph (|z|)^{2} 
\sum\limits_{n=0}^{\infty}\frac{|z|^{2n+1}}{e(n)}e^{-it(2\kappa
s(n+1)+\Phi-\alpha)}\frac{1}{s(n+1)},
\era
\eeq
where
\beq
\bra{llll}
q(n+1)=\frac{\frac{\Delta_- }{4\kappa}(e_{n+2}-e_n
)}{\sqrt{e_{n+1}+(\frac{\Delta_- }{4\kappa})^2 (e_{n+2}-e_n )^2 }},& q(0)=1\\
s(n+1)=\frac{1}{\sqrt{e_{n+1}+(\frac{\Delta_- }{4\kappa})^2 (e_{n+2}-e_n
)^2 }},& s(0)=1\\
\alpha = \arctan z ,& .
\era
\eeq
The result of the computation of $\langle\sigma_{3}\rangle $ shows that
the time dependance comes from the value $\langle\sigma_{3}\rangle
_{+-}+\langle\sigma_{3}\rangle_{-+}$ and we obtain
\beq
\langle\sigma_{3}\rangle _{+-}+\langle\sigma_{3}\rangle_{-+}=\aleph(|z|)^2 
\sum\limits_{n=0}^{\infty}\frac{|z|^{2n+1}}{e(n)}(\frac{\sin\theta\cos(t(2\kappa
s(n+1)+\Phi-\alpha))}{s(n+1)}).
\eeq
This expression have an oscillating behaviour that characterizes the
atomic inversion in the generalized JC model.
\section{Conclusion}
\hspace{.6cm}In this work, we developed the generalization of the
Jaynes-Cummings model where the radiatif field is replaced by generalized
harmonic oscillators. We shown the exact solvability of the model. The role
of the nonlinear dynamical superalgebra $u(1/1)$, governing the evolution of
the related model, is important in the construction of the supercoherent 
states over which we compute the energy, average number of photons and
the atomic inversion. These states has been constructed using the unitary
transformations expressed in terms of the generators of the $u(1/1)$
superalgebra. Furthermore, it is shown that the time dependent supercoherent
states have the advantage to obtain the Rabi oscillations.

Finally, we note that the importance of the generalized Jaynes-Cummings
model is due not only to its exact solvability but also arises from its quantum
effects such as revival of atomic inversion. In our opinion the generalization
and results presented here can be used to give a realistic description of the
nonlinear process of the interaction of an atom and radiation field. Clearly,
only a confrontation with experimental measures can valid the results of this
work. 
\section{Acknowledgements}  
\hspace{.6cm}The authors are thankful to Abdus Salam International Centre
of Theoretical Physics (AS-ICTP). M. D would like to thank Professor Yu. Lu
for his kind invitation to joint the condensed matter section of AS-ICTP. He is
also grateful to Professor V. Hussin for useful discussions during the
elaboration of this work and kind hospitality at CRM-Montr\'eal. J. D thanks
the financial support from AS-ICTP in the framework of the associate
program.

\end{document}